\documentclass[referee]{aa} 
\usepackage{graphicx}
\begin{document}
   \title{On the evolution of the cosmic-mass-density contrast and the cosmological constant}

   \author{D. PALLE
          }

   \institute{Zavod za teorijsku fiziku, Institut Rugjer Bo\v {s}kovi\' {c}\\
             Bijeni\v {c}ka cesta 54, HR-10002 Zagreb, CROATIA}

   \abstract{
We study the evolution of the cosmic mass density contrast beyond
the Robertson-Walker geometry including the small contribution of
acceleration. We derive a second-order evolution equation
for the density contrast within the spherical model 
for CDM collissionless fluid including the cosmological constant,
the expansion and the non-vanishing four-vector of acceleration.
While the mass-density is not seriously affected by acceleration,
the mass-density contrast changes its shape at smaller redshifts even
for a small amount of the acceleration parameter. This could help to resolve 
current controversial results in cosmology from measurements of WMAP, gravitational
lensing, XMM X-ray cluster or type Ia supernovae data, etc.
   \keywords{cosmology --
                structure formation --
                cosmological constant
		\\
{\bf Dedicated to the memory on my father Velimir Palle.}
               }
   }
   \titlerunning{On the evolution of the cosmic-mass-density contrast}
   \maketitle
%

\section{Motivation}

With more and more precise measurements in cosmology, theorists are confronted
with more and more questions with no reasonable answers.
However, there is good agreement on the value of the total mass density (flat Universe),
the Hubble constant, baryon density or the adiabatic spectrum of the initial mass-density
fluctuations. Einstein's gravity supplemented with the inflaton scalar is 
the key theoretical framework to understand the dynamics and kinematics of
the evolution of the Universe. Unfortunately, even the extreme freedom to model
a scalar potential or to add more inflaton scalars cannot explain large-angle
WMAP data or controversial and contradictional conclusions from data at lower scales from XMM X-ray clusters,
supernovae, CMBR, lensing etc. The introduction of dark matter and "dark energy"
plays the essential role in evading theoretical problems to fit the data.

Our attempt to understand the large and the small scale structure
of spacetime relies on the study how to solve or to avoid deficiencies
of the interactions that govern the large and the small scales
in the Universe: general relativity(GR) and local gauge interactions.
Trying to solve the common problem of the zero-distance singularity,
we concentrate on the study of the nonsingular Einstein-Cartan (EC)
cosmology (Palle, 1996a,1999,2001a,2002a) and the nonsingular SU(3) conformal local gauge
theory (Palle, 1996b,2000,2001b,2001c,2002b). 

Here we briefly exhibit our main results: (1) the minimal scale of the EC
cosmology is compatible with the weak interacion scale, (2) the global vorticity and 
acceleration of the Universe do not vanish and are correlated to the Hubble
expansion inducing also the normalization of the mean mass density and
the cosmological constant owing to the nontrivial bootstrap of the
number density in the energy-momentum to curvature and spin to torsion
relations, (3) one can derive primordial mass density fluctuations from
quantum fluctuations at the weak scale (no inflaton scalar included), 
(4) fermion and gauge boson particle spectrum can be 
explained within SU(3) conformal theory assuming ultraviolet(UV) finiteness and
the relation between boson and fermion mixing angles, (5) three fermion 
families appear as zero-, one- and two-node solutions of
Dyson-Schwinger UV-finite bootstrap equations and the UV finiteness
is a necessary condition for the existence of nontrivial
solutions (Maskawa and Nakajima, 1974a,1974b),
(6) heavy Majorana neutrinos are cold dark matter(CDM) particles  
(cosmologically stable because of the absence of the Higgs scalars
in the theory) that
can solve problems of diffuse photon background and the early
reionization of the Universe (Chen and Kamionkowski, 2003),
(7) broken lepton number and heavy Majorana neutrinos produce
leptogenesis which could induce baryogenesis in the expanding
Universe, 
(8) galactic annihilation of heavy neutrinos can explain
the ultrahigh-energy cosmic ray events,
(9) light-Majorana-neutrinos flavour mixing leads to the explanation
of solar, atmospheric, LSND, etc. neutrino anomalies,
(10) at the LHC we expect the confirmation of 
strong enhancement of the QCD amplitudes observed at TeVatron, Run I
beyond the scale of 200 GeV,
especially indicative at the analysis of the quotients of cross sections 
at two different center of mass energies, when large
systematic errors are removed; the enhancement is a natural
consequence of the UV-finiteness (noncontractibility of
the physical space).

In this paper we investigate the influence of the small
contribution of acceleration present in the geometry 
beyond that of the Robertson-Walker metric in the study
of the evolution of the density contrast including
the cosmological constant and CDM pressureless fluid.

\section{Deriving and solving the evolution equation}

It is very difficult to construct a cosmological model with the
geometry beyond that of the Robertson-Walker and
with the simple energy-momentum tensor of pressureless
fluid in the matter dominated era.
Let us remind on the relation in the EC cosmology
(nonsingular and causal in the G\"{o}del's sense)
assuming the spinning fermion matter dominance (Palle, 1996a):

\begin{equation}
\Sigma H_{\infty} = \frac{2}{\sqrt{3}} \omega_{\infty}, 
\end{equation}
\begin{eqnarray*}
\Omega_{\gamma, 0}={\cal O}(10^{-4}),\ \Omega_{tot, 0}={\cal O}(1),
\ \Omega_{\gamma, \infty} \equiv 0
\Longrightarrow \Omega_{tot, 0}\simeq \Omega_{tot, \infty} . 
\end{eqnarray*}

Some evidence for the existence of the vorticity of the Universe
gives us the possibility of estimating the acceleration parameter
(Palle, 1996a; Li, 1998)

\begin{equation}
\omega_{0} = {\cal O}(10^{-13} yr^{-1}) \Longrightarrow \Sigma = {\cal O}
(10^{-3}) .
\end{equation}

Inspecting the conservation equation for pressureless fluid
with the metric containing expansion and acceleration one gets

\begin{equation}
ds^{2} = dt^{2}-a^{2}(t)[(1-\Sigma)dr^{2}+r^{2}(d\theta^{2}
+sin^{2}\theta d\phi^{2})]-2\sqrt{\Sigma}a(t)dr dt, 
\end{equation}
\begin{equation}
a^{\mu}\equiv u^{\nu}\nabla_{\nu} u^{\mu},\ a^{\mu}a_{\mu}
= -\Sigma \frac{\dot{a}^{2}}{a^{2}}, \dot{a}\equiv \frac{da}{dt},
\end{equation}
\begin{eqnarray*}
u^{\mu}=four-vector\ of\ velocity;\ a^{\mu}=four-vector\ of
\ acceleration, 
\end{eqnarray*}
\begin{equation}
T^{\mu\nu}_{;\nu}=0 \Longrightarrow \rho_{m} (a) \propto a^{-3-\Sigma}.
\end{equation}

Thus, for a small acceleration parameter there is a negligible 
influence on the scaling of the mass density in
the matter dominated era.

Our attempt to treat density fluctuations assumes the standard 
scaling of the background $\rho_{m}(a)\propto a^{-3}$. 
We start with the equations of motion of particles (Weinberg, 1972):

\begin{equation}
\frac{d^{2}x^{i}}{dt^{2}} = -\Gamma^{i}_{\nu\lambda}\frac{dx^{\nu}}{dt}
\frac{dx^{\lambda}}{dt}+\Gamma^{0}_{\nu\lambda}\frac{dx^{\nu}}{dt}
\frac{dx^{\lambda}}{dt}\frac{dx^{i}}{dt} .
\end{equation}

Spherical symmetry and
a transition from the comoving to the physical radial coordinate leads to:

\begin{eqnarray}
\ddot{R}-\frac{\ddot{a}}{a}R+\sqrt{\Sigma}\frac{\dot{a}}{a} \hspace{60 mm}\nonumber \\
= \frac{\dot{a}}{a}[3\Sigma (\dot{R}-\frac{\dot{a}}{a}R)
-3\sqrt{\Sigma}(-1+\Sigma)(\dot{R}-\frac{\dot{a}}{a}R)^2
+(-1+\Sigma)^2(\dot{R}-\frac{\dot{a}}{a}R)^3], \\
R\equiv  r a(t),\ R=physical\ coordinate, 
\dot{R}\equiv \frac{dR}{dt}. \nonumber
\end{eqnarray}

Now we refer to and assume the spherical model where the spherical shells
do not cross each other during the evolution (Peebles, 1980; Padmanabhan, 1993)

\begin{eqnarray}
\dot{r}=0,
\end{eqnarray}
\begin{equation}
M=\frac{4\pi}{3}\rho_{b}(1+\delta_{i})R^{3}=const. ,
\end{equation}
\begin{eqnarray*}
H_{0}^{2} = \frac{8 \pi G_{N}}{3}\rho_{c,0}.
\end{eqnarray*}

The right-hand side of the Eq. (7) vanishes in the spherical model, thus 
we are left with

\begin{equation}
\ddot{R}+G_{N}MR^{-2}-\frac{8\pi G_{N}}{3}\Lambda R+\sqrt{\Sigma}
\frac{\dot{a}}{a} = 0,
\end{equation}
\begin{eqnarray*}
\Omega_{m}+\Omega_{\Lambda}=1,\ \frac{\dot{a}}{a}=
H_{0}[\Omega_{m}(R/r)^{-3}+\Omega_{\Lambda}]^{1/2},
\end{eqnarray*}
\begin{eqnarray*}
M=\frac{4\pi}{3}\Omega_{m}\rho_{c,0}R_{U}^{3}(1+\delta_{i}),
\end{eqnarray*}
\begin{eqnarray*}
R_{U}=H_{0}^{-1},\ \Lambda=\rho_{c,0}\Omega_{\Lambda}.
\end{eqnarray*}

One can easily perturb this equation acknowledging the easily
proved relations $\delta\dot{R}=(\delta R)^{.},\ 
\delta\ddot{R}=(\delta R)^{..}$:

\begin{eqnarray}
\ddot{\delta}&-&\frac{2}{3}\frac{\dot{\rho_{b}}}{\rho_{b}}\dot{\delta}
-\frac{1}{3}(\frac{\ddot{\rho_{b}}}{\rho_{b}}
-\frac{4}{3}\frac{\dot{\rho_{b}}^{2}}{\rho_{b}^{2}})\delta  \nonumber \\
&-&\frac{8\pi G_{N}}{3}(\rho_{b}+\Lambda)\delta
-\frac{3}{2}\sqrt{\Sigma}R^{-1}H_{0}\Omega_{m}a^{-3}
(\Omega_{m}a^{-3}+\Omega_{\Lambda})^{-1/2}\delta=0,
\end{eqnarray}
\begin{equation}
\delta \equiv \frac{\delta \rho}{\rho}.
\end{equation}

Writing it with more suitable variables, we get

\begin{eqnarray}
&&\delta^{\prime\prime}+\frac{3}{2}a^{-1}(\Omega_{m}a^{-3}+
2\Omega_{\Lambda})(\Omega_{m}a^{-3}+\Omega_{\Lambda})^{-1}
\delta^{\prime} \nonumber \\
&&-\frac{3}{2}a^{-5}\Omega_{m}(\Omega_{m}a^{-3}+\Omega_{\Lambda})^{-1}
\delta -\frac{3}{2}\Omega_{m}^{4/3}\Sigma_{0}^{1/2}a^{-6-1/4}
(\Omega_{m}a^{-3}+\Omega_{\Lambda})^{-3/2}\delta=0,
\end{eqnarray}
\begin{equation}
\delta^{\prime}\equiv \frac{d \delta}{d a},\ a(z)=1/(1+z).
\end{equation}

Assuming the varying acceleration parameter, from Eq. (1) one can 
conclude (Li, 1998) that

\begin{equation}
\omega \propto a^{-2},\ H \propto a^{-3/2},\ \Sigma \simeq \omega H^{-1}
\end{equation}
\begin{equation}
\Longrightarrow \Sigma = \Sigma_{0} a^{-1/2}.
\end{equation}

Then, the evolution equation contains one additional term

\begin{equation}
\delta^{\prime\prime} + ... 
-\frac{1}{4}\Sigma_{0}^{1/2}\Omega_{m}^{1/3}a^{-3-1/4}
(\Omega_{m}a^{-3}+\Omega_{\Lambda})^{-1/2}\delta=0.
\end{equation}

To solve equations (13) and (17) we have to fix initial conditions
for the very small cosmic scale factor a(z). Eq. (17) then looks as

\begin{equation}
\delta^{\prime\prime}+\frac{3}{2}a^{-1}\delta^{\prime}
-\frac{3}{2}a^{-2}\delta -\frac{3}{2}a^{-2}\beta \delta = 0,
\end{equation}
\begin{eqnarray*}
\beta&=&\frac{7}{6}\Sigma_{0}^{1/2}\Omega_{m}^{-1/6}a^{1/4},\ for\ 
\Sigma=\Sigma_{0}a^{-1/2}, \\
\beta&=&\Sigma_{0}^{1/2}\Omega_{m}^{-1/6}a^{1/2},\ for\ 
\Sigma=\Sigma_{0}=const.
\end{eqnarray*}

It can be approximately solved by the Ansatz for the small cosmic scale factor:

\begin{equation}
\delta \propto a^{\alpha},\ \alpha\simeq const.,
\end{equation}
\begin{eqnarray*}
\delta_{\pm}\propto  a^{\alpha_{\pm}},\ 
\alpha_{\pm} = -\frac{1}{4}\pm\frac{1}{2}\sqrt{\frac{1}{4}+6(1+\beta)}.
\end{eqnarray*}

From these solutions for the small scale factor we can obtain initial conditions
for a second-order evolution equation (17) in the
matter dominated era.

For vanishing acceleration, one gets growing and decaying modes
as for sub-horizon density perturbations (Kolb and Turner, 1990; Padmanabhan, 1993) because the
right-hand side of (7) in the derivation of the evolution equation can be neglected 
($R/d_{H}<<1$) in comparison with the left-hand side. In the spherical
model, the right-hand side of Eq.(7) vanishes exactly for super- or sub-horizon 
perturbations.

\begin{equation}
\Sigma=0 \Rightarrow\ \delta_{+} \propto a,\ 
\delta_{-} \propto a^{-3/2}.
\end{equation}

Anyhow, the physically relevant growing mode agrees for super- and sub-horizon
perturbations $\delta_{+} \propto a$ (Kolb and Turner, 1990).

We can now solve and compare the solutions of Eq. (17) for various parameters
and initial conditions (Adams-Bashforth integration method used):

\begin{equation}
(\frac{\delta T}{T})_{i}\simeq 1.1\times 10^{-5},\ 
h_{0}\simeq 0.71 \Rightarrow \delta_{i}\simeq 60 h_{0}^{2} (\frac{\delta T}{T})_{i}
= 3.33\times 10^{-4},
\end{equation}
\begin{eqnarray*}
z_{dec}\simeq 1150.
\end{eqnarray*}

The results are visualized in Figs. 1-3.

   \begin{figure}
   \centering
   \includegraphics[width=\textwidth]{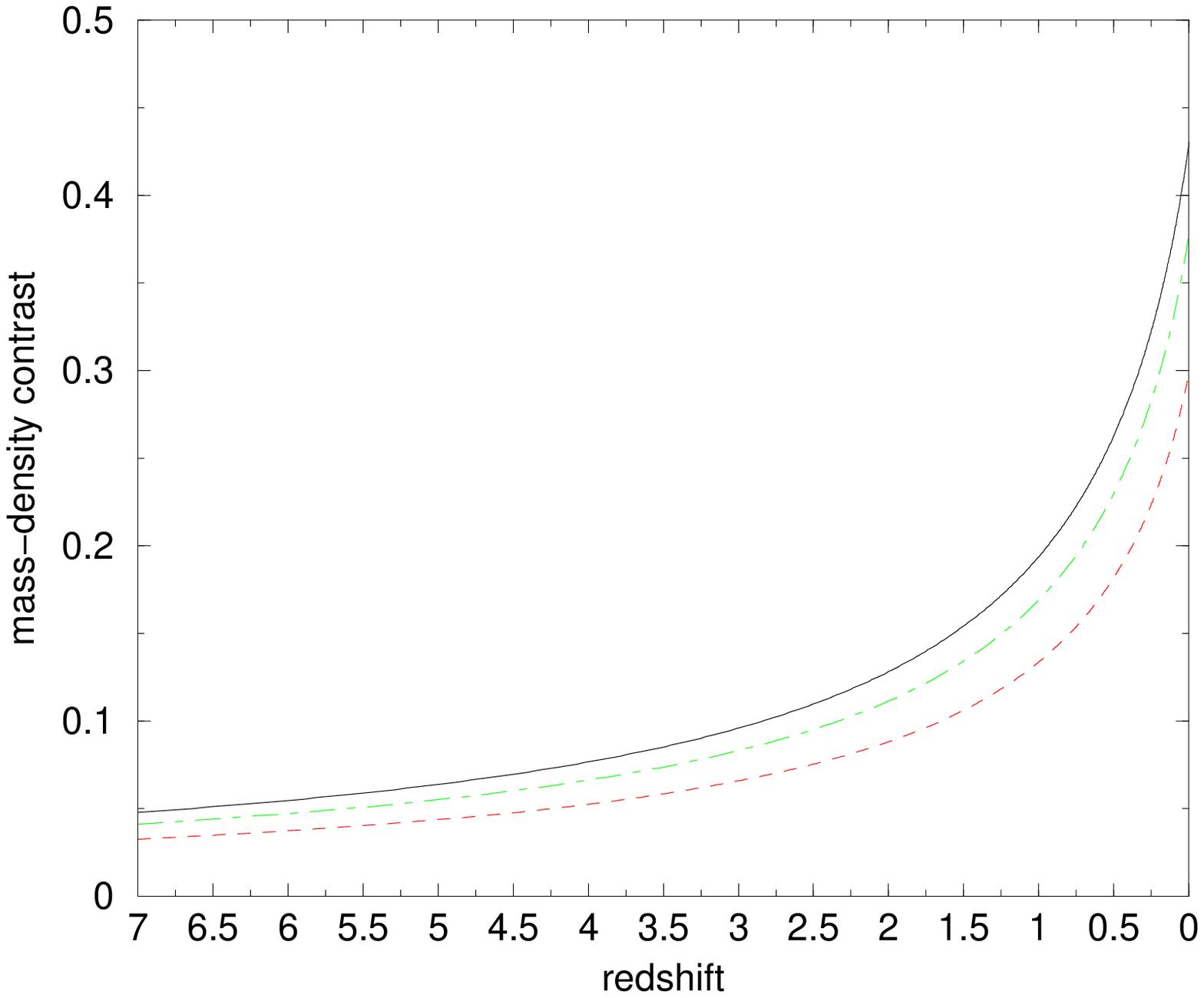}
     \caption{Mass-density contrast vs. redshift: $\delta_{i}=3.33\times 10^{-4},\   
     \Omega_{m}=2,\ \Sigma=0,\ 0.4\times 10^{-3},\ 0.4\times 10^{-3} a^{-1/2}$
     denoted, respectively, by solid, dashed and dot-dashed lines.}
   \end{figure}

   \begin{figure}
   \centering
   \includegraphics[width=\textwidth]{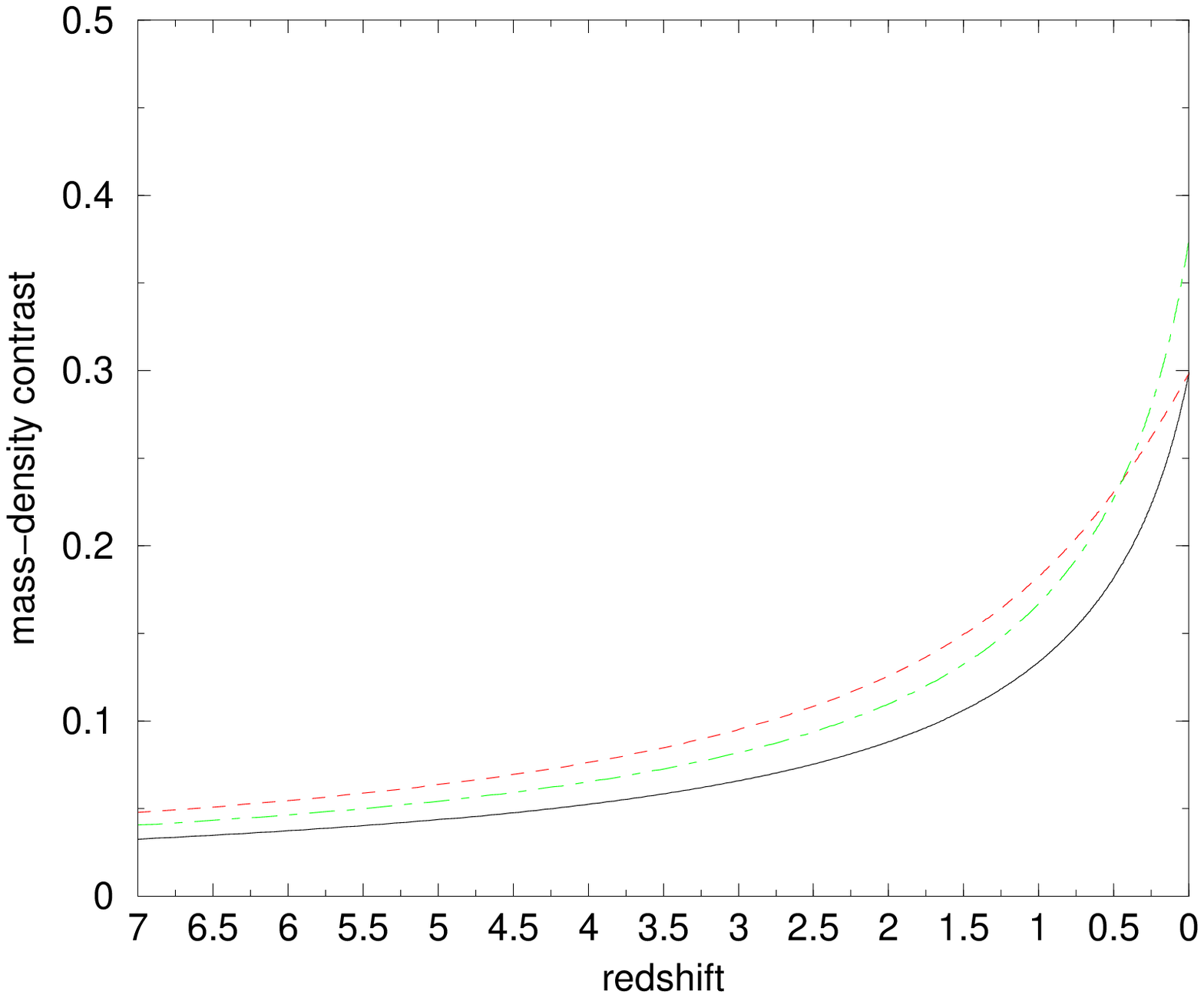}
      \caption{Mass-density contrast vs. redshift: $\delta_{i}=3.33\times 10^{-4};
      \ 3.33\times 10^{-4};\ 2.9\times 10^{-4},\ 
      \Omega_{m}=2;\ 0.3;\ 2;\ \Sigma=0.4\times 10^{-3};\ 0;\ 0.8\times 10^{-3}a^{-1/2}$
	       denoted, respectively, by solid, dashed and dot-dashed lines.}
   \end{figure}

   \begin{figure}
   \centering
   \includegraphics[width=\textwidth]{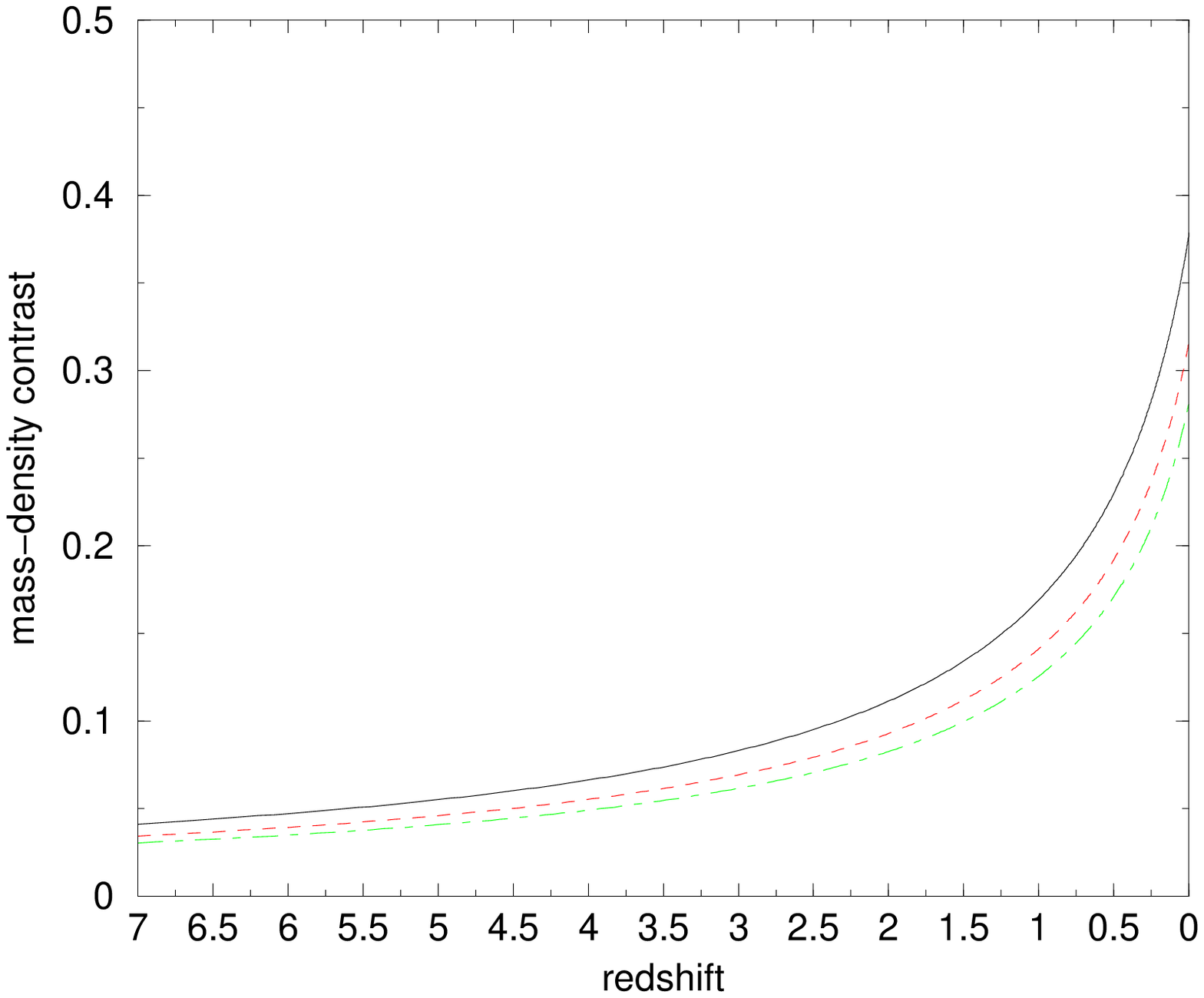}
      \caption{Mass-density contrast vs. redshift:
      $\delta_{i}=3.33\times 10^{-4};\ 3.33\times 10^{-4};\ 2.9\times 10^{-4},
      \ \Omega_{m}=2;\ 2;\ 2,\ \Sigma=0.4\times 10^{-3}a^{-1/2};
      \ 0.8\times 10^{-3};\ 10^{-3}$
      denoted, respectively, by solid, dashed and dot-dashed lines.}
   \end{figure}

\section{Discussion and conclusions}

From the preceding section we may conclude that the mass density contrast 
in the matter dominated era is substantially affected
by the small amount of acceleration beyond the Robertson-Walker geometry
although rendering the mass density almost unaffected.
Similar conclusions one may also expect for baryon matter, but then 
the collisions (Peebles, 1980, Padmanabhan, 1993)
should be also included in the evolution calculus.

It is evident that smaller initial density fluctuation can be compensated
by larger (but still small) acceleration parameter resulting in
similar behaviour of density contrast at smaller redshifts.
Such an adjustment can also be made for different mass-density 
models.

Some cosmic observables depend on both density and density contrast,
while others depend on only one of them. These can be a source of confusion
if the geometry of the Universe contains acceleration.
For example, the theoretical analysis of the CMBR implies
equations of coupled multicomponent fluid depending
on all relevant cosmic variables including density and
its contrast.
The small power at largest distances, as observed in the first
year data of WMAP, is probably due to the integrated Sachs-Wolfe
effect of the EC model with the negative cosmological
constant, while the small scale structure calculations should
be implemented with nonvanishing acceleration. 
On the other hand, the XMM X-ray cluster data analysts
claim large mass density, in apparent contradiction
with data sensitive to density contrasts.

To conclude, it is obvious that only a more general geometry
of spacetime with the inclusion of expansion, rotation,
acceleration, shear, torsion and 
definite EC cosmology prediction for the cosmological constant (Palle, 1996a)
can save us from
rather speculative considerations.
In addition, the SU(3) conformal gauge theory (Palle, 1996b) supplies
mass and spin of the matter content for the Universe: cold and hot dark matter as 
heavy and light neutrinos, respectively, baryons, charged leptons and the photon.
The very fine future measurements of gravity by the LATOR mission,
for example, at the solar system scale
could be a complementary way to find a value of the cosmological constant.
\newline

\hspace{40 mm} ***
\newline
I avoid referencing observational or experimental
work because of extremely overwhelming number of
papers with fantastic technological achievements and physical
results that could produce a large number of reference-list pages.
I apologize to all these astrophysicists and refer the reader 
to e-print archieves and web-sites of the corresponding projects.

\end{document}